\documentclass[aps,prl,floats,twocolumn,superscriptaddress,showpacs]{revtex4}

\usepackage[dvips]{epsfig}
\usepackage[dvips]{graphicx}
\usepackage{latexsym,amsmath}
\usepackage{boxedminipage}
\usepackage{color}

\newcommand{\be}{\begin{equation}}
\newcommand{\ee}{\end{equation}}
\newcommand{\bea}{\begin{eqnarray}}
\newcommand{\eea}{\end{eqnarray}}

\begin{document}

\title{Irrelevance of information outflow in opinion dynamics models}

\author{Claudio Castellano} 
\affiliation{CNR-ISC, UOS Sapienza \\ and Dipartimento di Fisica, ``Sapienza''
  Universit\`a di Roma, P.le A. Moro 2, I-00185 Roma, Italy}
\author{Romualdo Pastor-Satorras} 
\affiliation{Departament de F\'\i sica i Enginyeria Nuclear,
  Universitat Polit\`ecnica de Catalunya, Campus Nord B4, 08034
  Barcelona, Spain}

\date{\today}

\begin{abstract}
  The Sznajd model for opinion dynamics has attracted a large interest
  as a simple realization of the psychological principle of social
  validation.  As its most salient feature, it has been claimed that
  the Sznajd model is qualitatively different from other ordering
  processes, because it is the only one featuring outflow of
  information as opposed to inflow. We show that this claim is
  unfounded by presenting a generalized zero-temperature Glauber-type
  of dynamics which yields results indistinguishable from those of the
  Sznajd model. In one-dimension we also derive an exact expression
  for the exit probability of the Sznajd model, that turns out to
  coincide with the result of an analytical approach based on the
  Kirkwood approximation. This observation raises interesting
  questions about the applicability and limitations of this approach.
\end{abstract}


\maketitle

In the last decades statistical physics has crossed many boundaries
between different fields, becoming, with its methods and concepts, a
powerful tool for the investigation of a broad range of disciplines.
This process has been mutually beneficial, since the consideration of
problems far away from a purely physical motivation has greatly
broadened the kind of theoretical questions and conceptual challenges
statistical physics is called to tackle.  One of the settings in which
this cross-fertilization has been particularly fruitful is
opinion dynamics~\cite{Castellano09}, where the goal is to
understand how global consensus/understanding/agreement emerges
out of disorder, based on local interactions. In this field
many simple models akin to those of statistical physics have been
introduced, both by social scientists and
physicists~\cite{Clifford73,axelrod97,Galam82,Deffuant00,Krapivsky03},
leading to intense activity and remarkable results.  In this context,
the model introduced by Sznajd-Weron and Sznajd~\cite{Sznajd00},
commonly denoted as Sznajd model (SM), has enjoyed an exceptional
success as the first one encoding the principle of ``social
validation'', stating that the convincing power of an individual
is greatly enhanced if another individual supports the same
view.

The dynamics of the SM in one dimension is defined as follows
\cite{Sznajd-Weron:830585, Stauffer00}: Each site in a one dimensional
lattice is endowed with a binary variable (spin) $\sigma_i = \pm 1$.
At each time step a pair of neighboring sites is selected at random,
$i$ and $i+1$. If these individuals have the same opinion, $\sigma_i =
\sigma_{i+1} \equiv \sigma$, the opinion of all the neighbors of $i$
and $i+1$ changes to the common value $\sigma$; otherwise, nothing
happens~\footnote{A variant of the Sznajd model (``Sznajd B'' dynamics
  in Ref.~\cite{Castellano09} has been shown~\cite{Behera03} to be
  perfectly equivalent to voter dynamics with next-nearest neighbors
  interactions and hence not encoding any ``social validation''.}.
The process is iterated until, on a finite system, a final consensus
(all spins equal) is reached.  Generalizations to higher dimensions
have been introduced and are described below.  Typical quantities of
interest are the consensus (fixation) time $T(x,N)$, defined as the
time needed to reach the state with all spins equal for a system of
size $N$, starting from a configuration with a fraction $x$ of
positive spins, and the exit probability $E(x)$, defined as the
probability that the final state will be all $\sigma_i=+1$.

The Sznajd model is similar to other simple models for dynamics of
Ising spins in the absence of bulk noise, such as the voter model and
the zero temperature Glauber dynamics. However, much emphasis has been
put~\cite{Krupa05,Sznajd06,Sousa06} on the claim that SM is
fundamentally different because it is the only model where
``information flows out'' (i.e. spins propagate their state to their
neighbors) as opposed to other models where a central spin adapts
itself to the state of the surrounding ones ("information inflow").
This claim is mainly supported by the shape of $E(x)$ in one
dimension, which is linear for the Glauber zero temperature dynamics
(as well as for the voter model), while it is nontrivial for SM
\cite{Lambiotte08,Slanina08}. Also the consensus time $T(x,N)$ has a
dependence on $x$ for SM that is not found in other types of dynamics.

In this Letter we show this claim to be unfounded, by presenting two
clearly ``outflow'' and ``inflow'' dynamics, given by simple
extensions of the SM and the Glauber models, respectively, in which
the number of sites involved in a single spin update is a model
parameter. The analysis of these models allows us to check that the
postulated difference between ``inflow'' and ``outflow'' dynamics in
fact does not exist. In particular, we show that in one dimension the
exit probabilities and consensus times of both models are the same.
The consideration of the two dimensional and mean-field cases adds
additional strength to our result.  Additionally, we provide an
exact expression for the exit probability of SM in one dimension,
revealing that previous results based on a Kirkwood approximation are
also exact, due to some surprising cancellation of errors that remains
to be understood.

The models we consider are defined in one dimension as follows:

\textit{Sznajd Model of range $R$, $SM(R)$}: At each time step, a pair
of nearest neighbor sites, $i, i+1$, is chosen at random. If they
share the same state, $\sigma_{i} = \sigma_{i+1} \equiv \sigma$, then
the $2R$ neighbors, to left and right, respectively, change their
value to $\sigma$, i.e. $\sigma_j \to \sigma$, for $j \in [i-R, i-1]
\cup [i+2, i+1+R]$. Otherwise, nothing happens.  In this ``outflow''
dynamics, the opinion of two adjacent equal spins thus extends to all
their $2R$ neighbors, the case $R=1$ corresponding to the standard SM.

\textit{Zero temperature Glauber dynamics of range $R$, $G(R)$}: The
elementary step consists in randomly selecting a site
$i$ and evaluating the local field given by the sum of the $2R$ spins
in the interval $[i-R,i-1] \cup [i+1,i+1+R]$.  If the local field is
positive or negative, the variable $\sigma_i$ aligns with it.
Otherwise the spin is randomly set to $\pm 1$ with
probability $1/2$.  For any $R$ the dynamics is obviously of
``inflow'' type, as the central spin is affected by the state of
surrounding spins.  The case $G(1)$ coincides with
the usual zero-temperature Glauber dynamics.

\begin{figure}
  \begin{center}
    \includegraphics*[height=5cm]{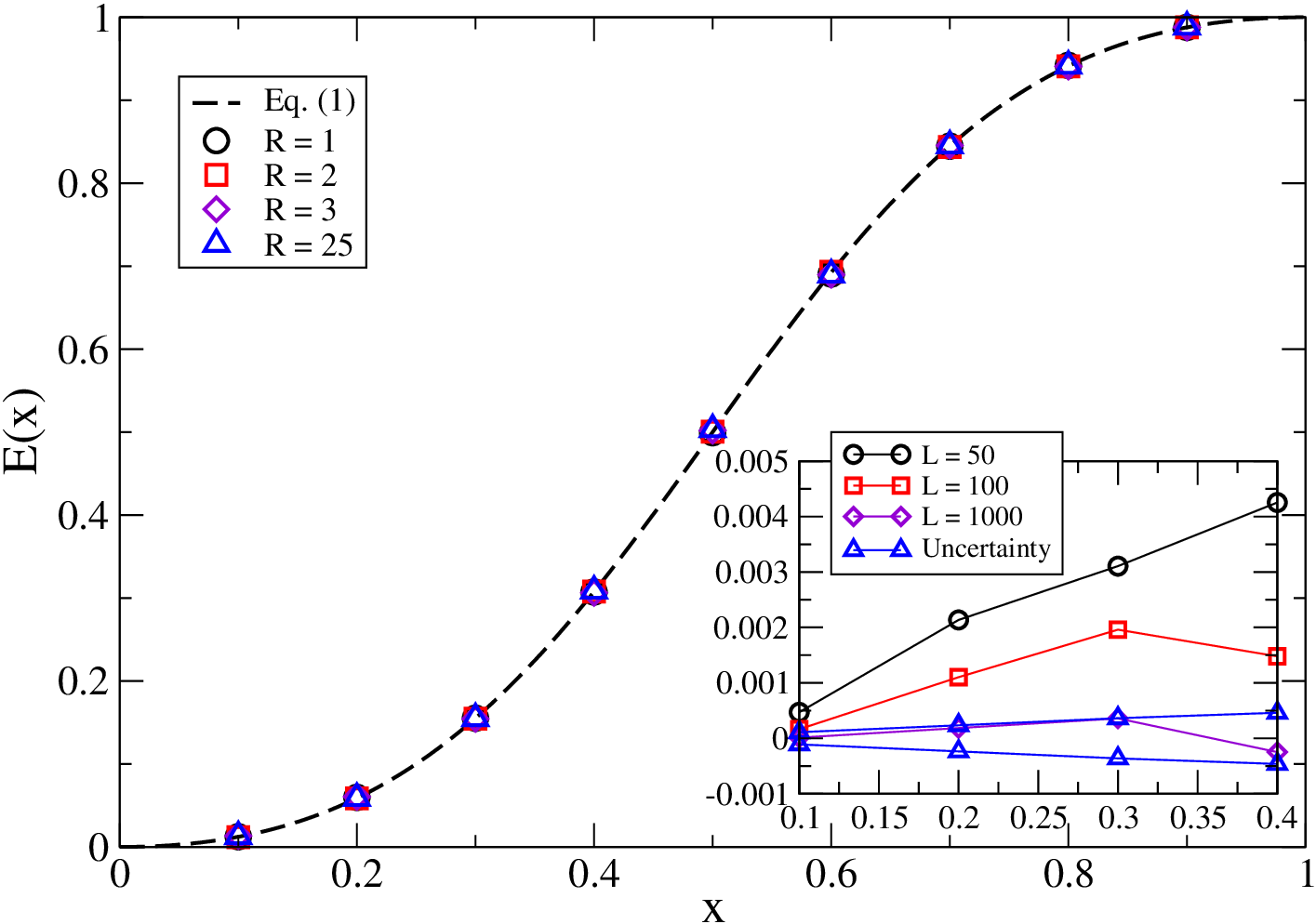}
  \end{center}
  \caption{Main: Exit probability for the $SM(R)$ on a one dimensional
    lattice of size $L=1000$.  The full line corresponds to the
    analytical prediction, Eq.~\eqref{cooleq}.  The number of
    realizations of the dynamics is $M=10^5$.  Inset: Difference
    between the numerical results for $SM(1)$ and the theoretical as
    $L$ changes. The uncertainty due to sampling error for $L=1000$ is
    given by $\pm \sqrt(E(x)[1-E(x)]/M)$, where the number of
    realizations of the stochastic process is $M = 10^6$.}
\label{fig1}
\end{figure}

Let us consider uncorrelated initial conditions in which each vertex
has a probability $x$ to be in the $+1$ state, and correspondingly a
probability $1-x$ to be in state $-1$.  As in other ordering processes
of this kind, the evolution in both $SM(R)$ and $G(R)$ models proceeds
in two separate stages.  Initially, homogeneous domains of spins up or
down quickly form at small scale. This stage lasts for a time of
the order of a few Monte Carlo steps per spin.  Later on, domain
boundaries diffuse around and annihilate upon encounter, leading to
larger and larger domains and eventually to consensus. The duration of
this second stage grows with the system size as $L^2$.  While in the
first stage the dynamics depends on the model's microscopic details
and the magnetization is not conserved, the second regime is very
similar for voter, generalized SM or generalized Glauber models, with
marginal variations due only to the details of the annihilation
process.  In this stage, the diffusion-annihilation boundary dynamics
leads to the conservation of the average magnetization.

We first study the behavior of the $SM(R)$ model, plotting in
Fig.~\ref{fig1} the exit probability for this model, computed
numerically for different values of $R$. Remarkably, $E(x)$ turns out
to be completely independent of the range of the interaction $R$.  By
taking advantage of this independence of $R$, we can derive the exact
form of the exit probability, which is very easy to compute for $R \ge
(L-2)/2$.  In such a case, the diffusive regime is absent and the
system becomes fully ordered after the first successful microscopic
update.  The dynamics proceeds by choosing at random $2$ consecutive
sites, that will be both in state $+1$ with probability $x^2$, in
state $-1$ with probability $(1-x)^2$, and in a mixed state with
probability $1 -x^2 -(1-x)^2$. The exit probability is given by the
probability that a pair of sites in state $+1$ is chosen before any
pair of sites in state $-1$. So, we can write
\begin{equation}
  \label{cooleq}
  E(x) = x^2 \sum_{n=0}^\infty [1 -x^2 -(1-x)^2]^n = \frac{x^2}{x^2
    +(1-x)^2}. 
\end{equation}
Another way to derive Eq.~(\ref{cooleq}), valid for smaller values of
$R$, is as follows: In the initial stage each successful update will
give rise to a domain of $2+2R$ equal sites, so that in a time of
order unity the system will be roughly subdivided into $L/(2 +2R)$
domains of size of order $2+2R$.  At the end of this
stage~\footnote{Note that the two types of dynamics are not sharply
  separated in time, but they are effectively independent.}, the
density of $+1$ spins will be $x'=1 \times x^2/[x^2+(1-x)^2] + 0
\times (1-x)^2/[x^2+(1-x)^2]$.  In the ensuing second stage the
conservation of magnetization implies that the exit probability is
$E(x')=x'$, independent of domain size, yielding again
Eq.~\eqref{cooleq}.

Fig.~\ref{fig1} shows that Eq.~(\ref{cooleq}) provides a very
accurate description of the exit probability of the generalized SM.
The inset of the figure proves moreover that the small deviations of
the numerical results for $E(x)$ around the theoretical value can be
fully ascribed to fluctuations around the expected value due to the
finite number of realizations of the process.  This
confirms that Eq.~(\ref{cooleq}) is the {\em exact } solution of the
exit probability for the SM.

It is crucial to remark that Eq.~(\ref{cooleq}) coincides with the
expression for the exit probability of SM calculated by solving
analytically the hierarchy of equations for multi-spin correlation
functions within a Kirkwood approximation decoupling
scheme~\cite{Lambiotte08,Slanina08}.  In this case then, the Kirkwood
approximation turns out to provide an exact solution for the SM model.
This is indeed a striking result, since numerical tests show that the
assumptions made in the Kirkwood approximation are largely violated
during the dynamics.  

\begin{figure}
  \begin{center}
    \includegraphics*[height=5cm]{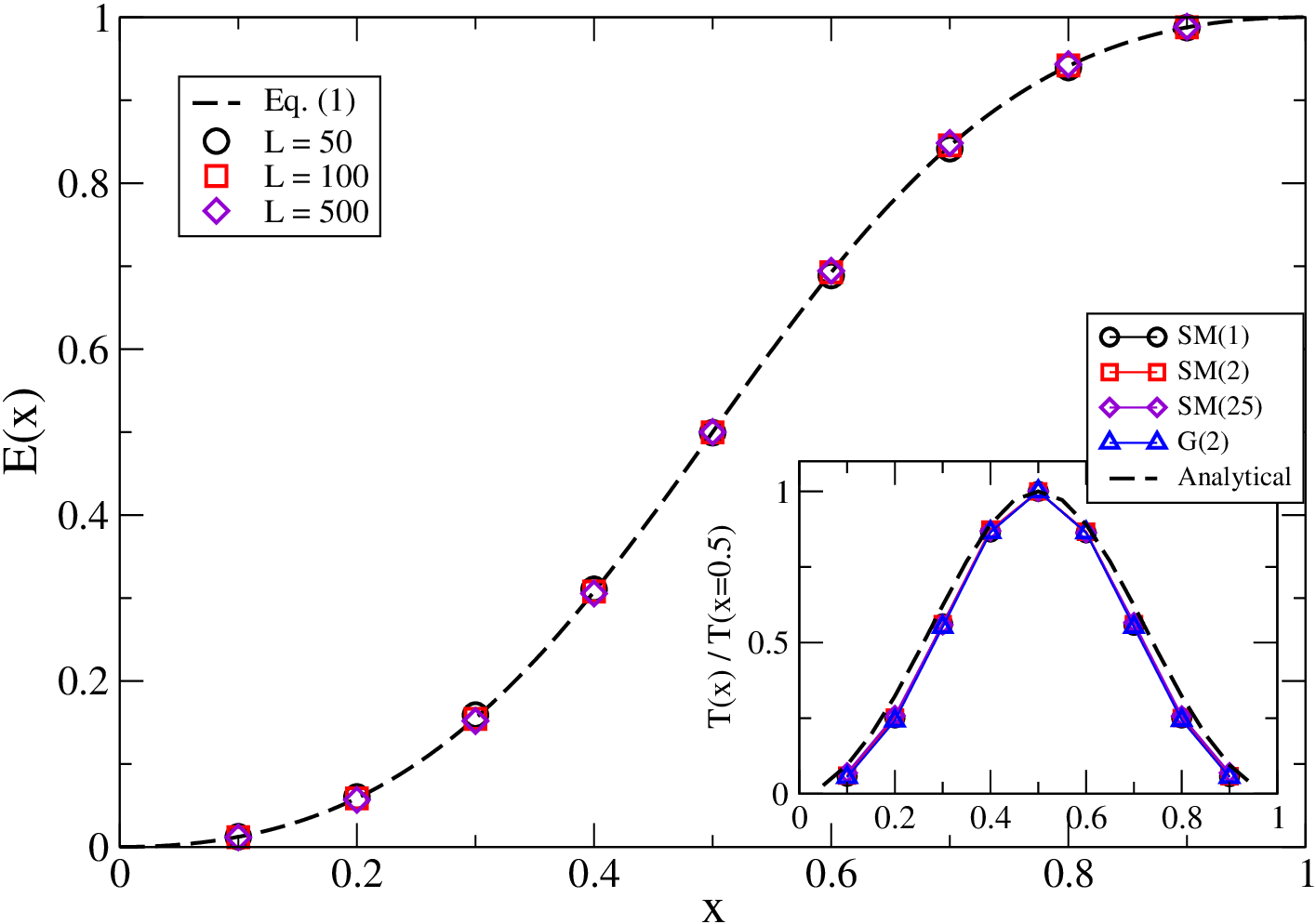}
  \end{center}
  \caption{Exit probability for the $G(2)$ dynamics
    in one dimensional lattices of increasing size.  The full
    line corresponds to the analytical prediction for the SM,
    Eq.~\eqref{cooleq}.  The number of realizations of the dynamics is
    $10^6$.  Inset: Numerical values of the consensus time $T(x)$
    rescaled by its maximum value $T(x=0.5)$, for $G(2)$ and $SM(R)$,
    compared with the analytical prediction (see text). Lattice size
    $L=500$.}
\label{fig2}
\end{figure}

Turning now to the generalized Glauber dynamics $G(R)$, numerical
simulations for $R=2$ (Fig.~\ref{fig2}) show that also in this case
$E(x)$ is in excellent agreement with Eq.~(\ref{cooleq}).  Hence the
exit probability of $SM(R)$ and of the $G(2)$ model are
indistinguishable.  The closeness of the two models is further
confirmed by the inset of Fig.~\ref{fig2}, where the consensus time
$T(x)$ (divided by $T(x=0.5)$ to factor out trivial temporal
rescalings) is reported: The time needed to reach the final consensus
is the same for both $SM(R)$ and $G(R=2)$ models. Fig.~\ref{fig2}
provides further evidence of the independence of the generalized SM
with $R$ and allows to conclude that direction of ``information flow''
is irrelevant: The behavior of Sznajd model with ``outflow'' dynamics
coincides with the behavior of $G(R=2)$ model, based on ``inflow''.

The dynamical division in two stages, illustrated above to derive the
exit probability, is useful also to obtain an analytical estimate of
the time $T(x)$ to reach consensus for the SM.  As described above,
in a time of order unity the density of $+1$ spins reaches its
asymptotic value $x' = x^2/[x^2+(1-x)^2]$, where $x$ is the
magnetization in the initial state. The ensuing evolution is
essentially the same followed by the voter model, for which the
consensus time is known and whose dependence on the initial density of
up spins $x'$ is $T \propto -[x' \ln x' + (1-x')
\ln(1-x')]$. Expressing $x'$ in terms of the initial value $x$ one
obtains an analytical formula for the consensus time of the SM.  The
comparison with numerics (Fig.~\ref{fig2}, inset) is rather good, the
discrepancy observed being probably ascribable to the slightly
different behavior of the models when two boundaries are one site far
apart. While in the voter dynamics they have equal probability to
collide or to go to distance 2, they deterministically collide in
SM.

The strong relationship between the $G(2)$ and SM is not
limited to one-dimensional systems.  Let us consider a random neighbor
topology, i.e. a fully connected system where the interaction occurs
with neighbors chosen randomly at each time step.  Slanina and
Lavicka~\cite{Slanina08} have analyzed the standard Sznajd dynamics
in this case, characterized by the transition rates
%($x=(1+m)/2$ is the density
%of up spins)
%
\begin{subequations}
\label{Rates1}
\bea
\textrm{Prob}[x \to x+1/N] &=& x^2(1-x), \\
\textrm{Prob}[x \to x-1/N] &=& x  (1-x)^2,
\eea
\end{subequations}
where $N$ is the system size.  For the generalized Glauber $G(2)$
dynamics the rates can be also easily worked out
\begin{subequations}
\label{Rates}
\bea
\textrm{Prob}[x \to x+1/N] &=& x^2(1-x)   (3 - 2 x), \\
\textrm{Prob}[x \to x-1/N] &=& x  (1-x)^2 (1 + 2 x).
\eea
\end{subequations}
The only variation is given by correcting factors which are smooth
and positive, thus implying that no basic feature of the dynamics will
change.  In particular, following the inverse Fokker-Planck formalism
\cite{Gardinerbook}, it is possible to show that the exit probability
takes in both cases the form of a Heaviside step function,
$E(x)=\Theta(x-0.5)$ for $N \to \infty$, as expected due to the presence
of an imbalance between the rates in Eqs.~(\ref{Rates1}) and~(\ref{Rates}).
\begin{figure}
  \begin{center}
    \includegraphics*[height=5cm]{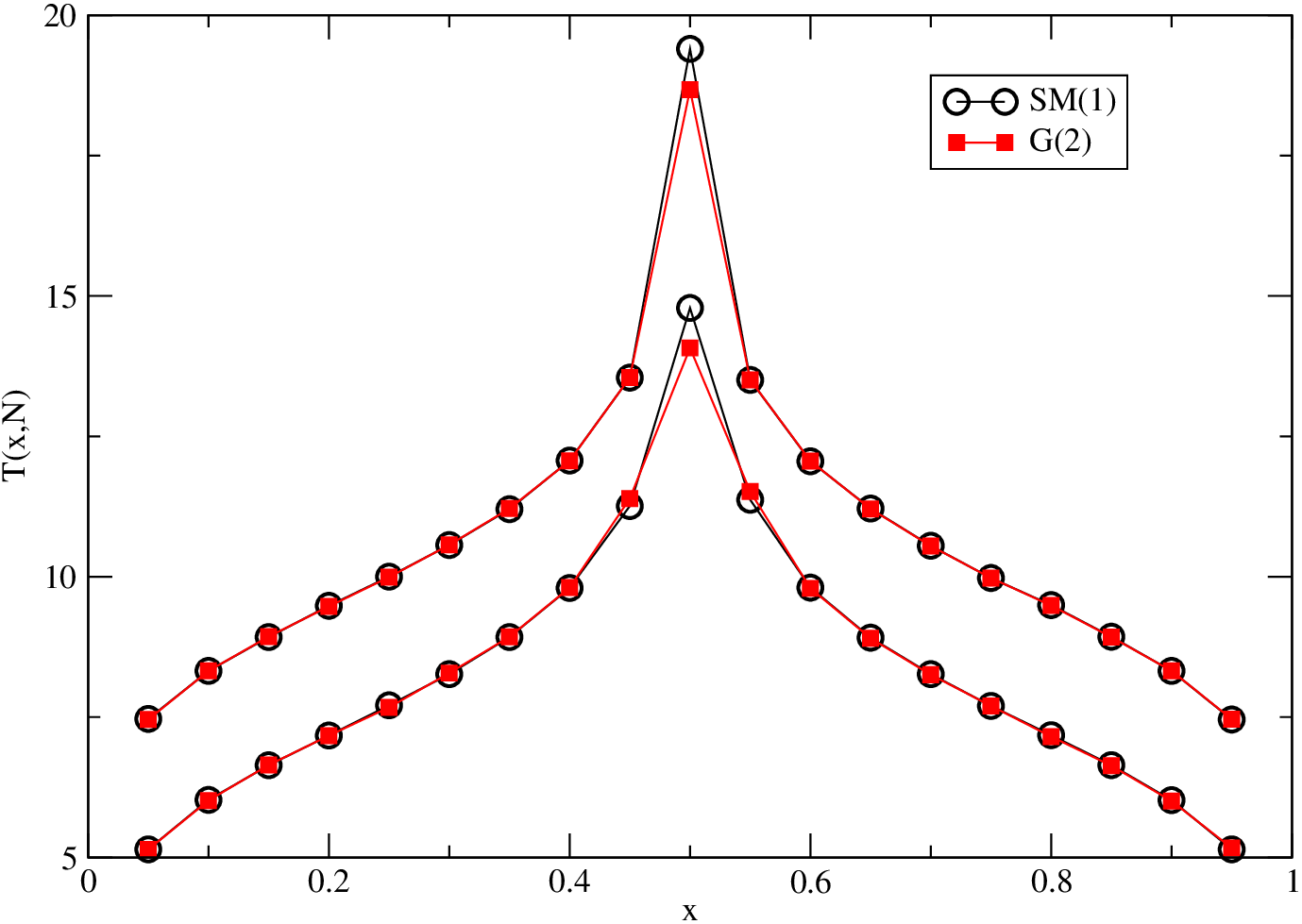}
  \end{center}
  \caption{Consensus time $T(x,N)$ as a function of the initial
    fraction $x$ of $+1$ spins, for $SM(1)$ and $G(2)$ models on a
    random neighbor topology.  Upper curves are for size
    $N=10^4$. Lower curves are for size $N=10^3$.}
\label{meanfield}
\end{figure}
Concerning the consensus time, we report the results of computer
simulations in Fig.~\ref{meanfield}, which prove the
equivalence between the SM and the $G(2)$ models at the mean-field
level.

In finite dimensions larger than one, many possible ways to define
SM have been introduced~\cite{Stauffer00,Krupa05,Kondrat08}.
Similarly, there are various possibilities to define the Glauber
dynamics for generic $R$.  We select the following ones. For the
$G(R)$ model, the local field for a site $(i,j)$ is given by the sum
of all spins up to the $R$-nearest neighbors. In particular for $R=2$
in $d=2$ the local field is given by the sum of the eight spins
surrounding $(i,j)$ and forming together a square of side $3$.  For
the SM, on the other hand, we consider two variants.  In SM-I dynamics
a bond is randomly chosen, either along the vertical or horizontal
direction, and if the sites at the extremes of the bond are equal, all
the 6 nearest neighbors of both sites are updated accordingly. In
SM-II, we select a plaquette of four sites and if they are in same
state, the 8 nearest neighbors are made equal.

The probability $E(x)$ to end up with all $+1$ spins is for all
variants of SM given (in the large size limit) by a step function
$E(x)= \Theta(x-0.5)$~\cite{Stauffer00}.  As can be expected based on
the fact that the dynamics is driven by curvature~\cite{Bray94}, the
same occurs for the $G(2)$ model, provided no freezing in a striped
configuration occurs~\cite{Spirin01}.  This phenomenon, which affects
asymptotically $G(1)$ dynamics in $d=2$ with a finite
probability~\cite{Spirin01} (and clearly affects $G(2)$ dynamics as
well), is present also in the evolution of the SM.  In this
case, straight stripes along one direction in a two-dimensional
lattice are not fully stable, given the intrinsic destabilizing
mechanism present in Sznajd dynamics at microscopic scales.
Nevertheless stripes do often form during the evolution and they
persist for very long times.
\begin{figure}
  \begin{center}
    \includegraphics*[height=5cm]{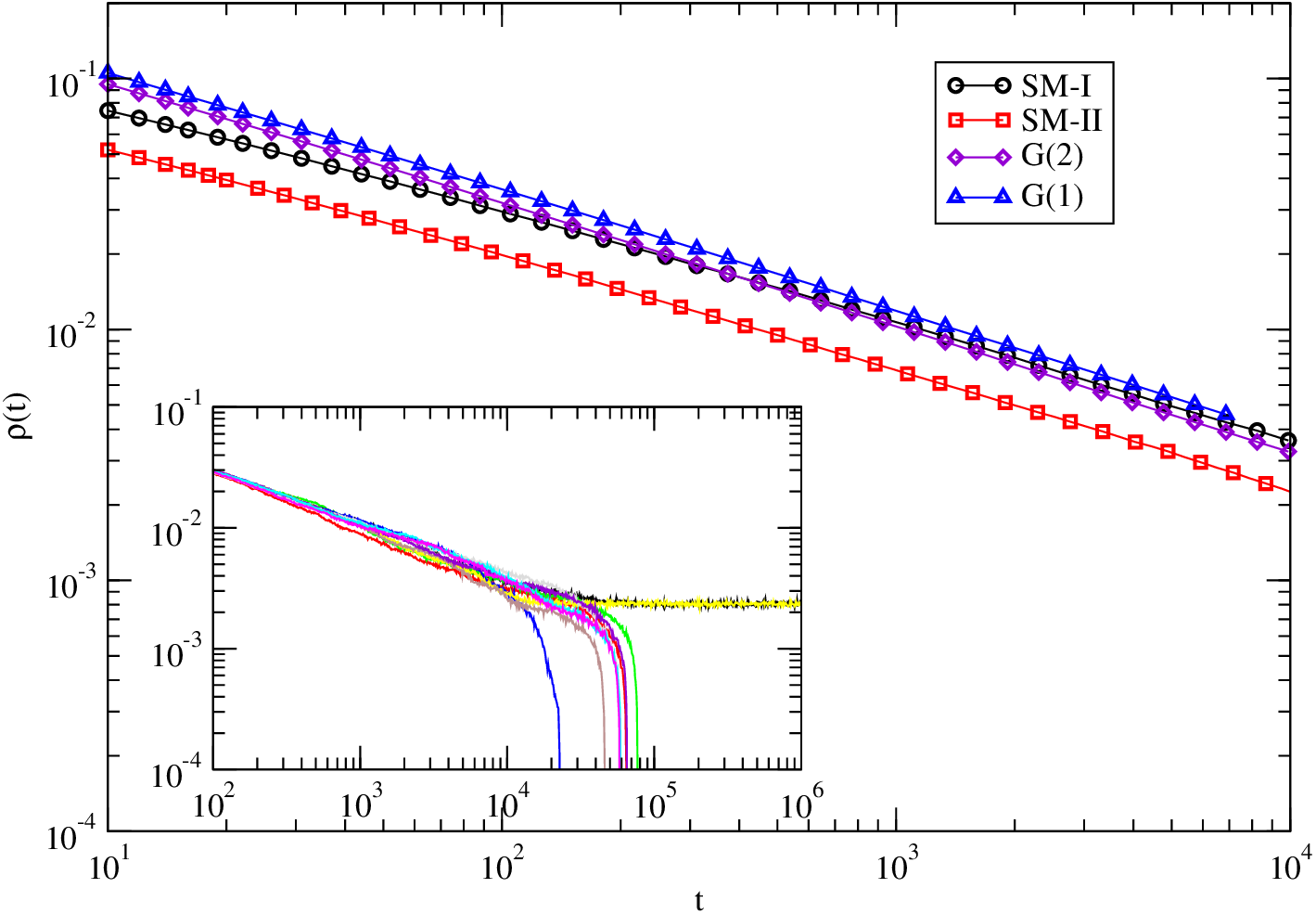}
  \end{center}
  \caption{Density $\rho(t)$ in $d=2$ for SM-I, SM-II, $G(1)$, and
    $G(2)$ dynamics.  System size $L=10000$.  Inset: $\rho(t)$ for
    several runs on a SM-I system of size $L=1000$, illustrating
    the stripe phenomenon.}
\label{2d}
\end{figure}
The presence of stable or long lived metastable striped states makes a
comparison between the consensus time $T(x,N)$ in SM and Glauber
$G(R)$ models impossible.  A quantity allowing a better analysis of
the ordering behavior of two-dimensional systems is the fraction
$\rho(t)$ of nearest neighbor pairs that are in opposite states.
Figure~\ref{2d} shows that for Sznajd and generalized Glauber dynamics
(with both $R=1$ or $R=2$) the evolution is the same, apart from
irrelevant transients and global temporal scales: The density $\rho$
decreases as $t^{-1/2}$, the signature of curvature-driven coarsening
dynamics~\cite{Bray94}.  On the other hand, the plateaus exhibited in
some realizations of Sznajd dynamics for long times (Fig.~\ref{2d},
inset) indicate the effective presence of long-lived metastable
states.  The perfect analogy between Sznajd and Glauber dynamics goes
beyond the decay of $\rho(t)$. The scaling functions for the two-point
correlation function $C(r,t)$ (not shown) are virtually the same.

In summary, we have shown that the behavior of Sznajd model for
opinion dynamics has no feature that distinguishes it from a
generalized zero temperature Glauber dynamics for Ising spins.
While in dimension $d>1$ this could be
expected on the basis of general considerations on coarsening systems,
in $d=1$ this result is highly nontrivial.  In one-dimensional
systems, the standard Sznajd dynamics actually differs from the usual
zero-temperature Glauber dynamics, as it has been extensively reported
in the literature.  However, when the range of the interactions is
extended to $R=2$, the generalized Glauber
dynamics is indistinguishable from Sznajd.  The conclusion is that
``outflow'' dynamics is not qualitatively different from
``inflow'' dynamics.
A possible objection to this conclusion is that inflow and outflow dynamics 
are actually different because $SM(R)$ does not
depend on $R$, while $G(R)$ dynamics does. This argument is rebutted
by considering another extension of SM, in which the number of equal
spins needed to convince neighbors is a parameter $q$. Numerical and
analytical arguments, to be reported elsewhere~\cite{Pastor10}, show
that $q$ strongly affects the dynamics and that such a generalized
``outflow'' dynamics gives results very close to those of the $G(R)$
``inflow'' model with $R=q$.  While studying the equivalence of $SM(R)$
and $G(2)$ we have derived an exact formula for the exit probability
of SM, that turns out to coincide with the one obtained by
using a Kirkwood approximation. Notice that, on the contrary,
Kirkwood approximation fails for the $G(2)$ dynamics~\cite{Pastor10}.
These findings call for
additional research to understand when the Kirkwood approximation
works, when it fails, and how it can be systematically improved.

\begin{acknowledgments}
  R.P.-S. acknowledges financial support from the Spanish MEC (FEDER),
  under projects No. FIS2007-66485-C02-01 and FIS2010-21781-C02-01;
  ICREA Academia, funded by the Generalitat de Catalunya; and the
  Junta de Andaluc\'{i}a, under project No. P09-FQM4682.  We thank
  S. Redner for helpful comments and discussions.
\end{acknowledgments}

%\bibliography{sznajd}

\end{document}